\documentclass[aps,prb,preprint,showpacs,superscriptaddress,floatfix]{revtex4}
\usepackage{graphics}
\usepackage{epsfig}
\usepackage{amsmath}
\usepackage{amssymb}
\usepackage{calc}
\newcommand{\beq}{\begin{equation}}
\newcommand{\eeq}{\end{equation}}
\newcommand{\bea}{\begin{eqnarray}}
\newcommand{\eea}{\end{eqnarray}}

\begin{document}
\title{Qubits from extra dimensions
}
\author{P\'eter L\'evay}
\affiliation{Department of Theoretical Physics, Institute of
Physics, Budapest University of Technology, H-1521 Budapest,
Hungary}
\date{\today}
\begin{abstract}
We link the recently discovered black hole-qubit correspondence to
the structure of extra dimensions. In particular we show that for
toroidal compactifications of type IIB string theory simple qubit
systems arise naturally from the geometrical data of the tori
parametrized by the moduli. We also generalize the recently
suggested idea of the attractor mechanism as a distillation
procedure of GHZ-like entangled states on the event horizon, to
moduli stabilization for flux attractors in F-theory
compactifications on elliptically fibered Calabi-Yau four-folds.
Finally using a simple example we show that the natural arena for
qubits to show up is an embedded one within the realm of
fermionic entanglement of quantum systems with indistinguishable
constituents.

\end{abstract}
 \maketitle{}

\section{Introduction}

In a remarkable paper Borsten et.al.\cite{Borsten} suggested that
wrapped branes can be used to realize qubits, the basic building
blocks used in quantum information. Based on the findings of that
paper it is natural to expect that such brane configurations
wrapped on different cycles of the manifold of extra dimensions
should be capable of accounting for the surprising findings of the
so called black hole qubit correspondence initiated in a series of
papers\cite{Duff,KL,Levay1} (for a review see the paper of Borsten
et.al\cite{review}). The aim of the present paper is to show that
by simply reinterpreting
some of the well-known results of toroidal compactification of
type IIB string theory in a quantum information theoretic fashion
this expectation can indeed be justified.
In particular we identify the Hilbert space giving home to the
qubits inside the cohomology of the extra dimensions, establishing
for the catchy phrase "to wrap or not to wrap, that is the
qubit"\cite{Borsten} a mathematical meaning, an issue left unclear
by Ref.1.

The black hole qubit correspondence is based on the observation
that the macroscopic Bekenstein-Hawking entropy formulas of
certain $4$ and $5$ dimensional black hole solutions of
supergravity models arising from compactifications of string and
$M$-theory happen to coincide with the ones of multipartite
entanglement measures used in the theory of quantum
entanglement\cite{Duff,KL,e7}. Though at first this observation
was merely regarded as an intriguing mathematical coincidence
however, it was soon realized that it can be quite useful on both
sides of the correspondence in a much wider context. In particular
we have learnt how to classify certain types of black hole
solutions using different classes of entanglement\cite{KL}, and
more importantly using the input provided by string theory we have
also seen how to obtain a {\it complete} solution to the
classification problem\cite{Verstraete} of entanglement types of
four-qubits using different classes of black hole
solutions\cite{Levay4qbit,
Duff4qbit}. The classification problem
under stochastic local operation and classical communication
\cite{Dur} (SLOCC) of entanglement classes for three qubits has
been revisited, and recovered in an elegant manner using
techniques originally developed within the realm of
 the supergravity
literature\cite{3qbitborsten}. More recently a classification
scheme for two-center black hole charge configurations for the
$stu$, $st^2$ and $t^3$ models based on the structure of
four-qubit SLOCC invariants and elliptic curves has been
proposed\cite{elliplevay,ferrorbit}. The structure of black hole
entropy formulas also inspired the construction of new and useful
tripartite measures for electron correlation and more generally
for quantum systems with both indistinguishable and
distinguishable constituents\cite{VL}. Moreover, using the input
coming from string theory it was shown that for such simple
quantum systems the SLOCC classification problem of entanglement
classes can be solved\cite{VL}.

Apart from issues concerning entanglement classes and their
associated entanglement measures, the black hole-qubit
correspondence also turned out to provide additional insight into
issues of dynamics of entangled systems. In particular it has been
shown that the well-known attractor mechanism\cite{attractors} of
moduli stabilization can be reinterpreted in the language of
quantum information as a distillation procedure of highly
entangled charge states on the event
horizon\cite{Levay1,Levayattr}. It was also realized that quantum
error correcting codes can be used to serve as a quantum
information theoretic framework for characterizing the properties
of the BPS and non-BPS attractors\cite{Levayattr,Frascati}.

What is the mathematical origin of the black hole-qubit
correspondence? Apart from arguments\cite{Duff,KL,Levay1,review}
based on the realization that on both sides of the correspondence
similar symmetry structures are present, none of these studies
have addressed the important question where are these qubits
reside, how the Hilbert spaces for the analogues of the usual
multipartite systems of quantum information are constructed. In
this paper we would like to make a step in the direction of
clarifying this important issue.

The crucial observation is the fact that  the various aspects of
supergravity models amenable to a quantum information theoretical
interpretation can all be obtained from {\it toroidal}
compactifications of type IIA, IIB or M-theory. Hence it is
natural to link the occurrence of qubits and qutrits in these $4$
and $5$ dimensional scenarios to the geometric data of tori i.e.
to the extra dimensions. 

In this paper we will concentrate merely
on qubits and work in the type IIB duality frame.
In Section II. as a warm up excercise, we show how deformed tori
give rise to a parametrized family of one-qubit systems. In
Section III. we analyse the archetypical example of the black hole
qubit correspondence-the $stu$ model\cite{Behrndt}. Coming from
compactification on a six torus $T^6$ in the type IIB duality
frame this model is featuring three-qubit systems. However, unlike
our warm up exercise this case already featuring entanglement,
namely the tripartite one. The attractor mechanism as a
distillation procedure\cite{Levayattr} is shown to arise naturally
in this picture.  In Section IV. we generalize our constructions
to flux attractors\cite{Kalloshnew}. We show that the idea of
distillation works nicely within the context of F-theory
compactifications on elliptically fibered Calabi-Yau four-folds
too. Here the toroidal case gives rise to four-qubit systems. As
an explicit example we revisit and reinterpret the solution found
by Larsen and O'Connell\cite{OConnell} in the language of
four-qubit entangled systems. In section V. we emphasize that our
simple qubit systems associated with the geometric data of extra
dimensions (tori) are giving examples to entanglement between
subsystems with {\it distinguishable} constituents. However, by
studying a simple example we show that, in the stringy context the
natural arena where these very special entangled systems live is
really the realm of fermionic entanglement\cite{eckert,ghirardi} of
subsystems with {\it indistinguishable} parts. The notion "fermionic
entanglement" is simply associated with the structure of the
cohomology of $p$-forms related to $p$-branes. Our conclusions and
some comments are left for Section VI.

\section{One-qubit systems from deformed tori}

Let us consider a torus $T^2$ with its complex structure
deformations labelled by \beq\tau\equiv x-iy\qquad
y>0.\label{fura}\eeq Here our choice for $\tau$ to have a negative
imaginary part is dictated by the conventions used in the
supergravity literature\cite{Ferrarastu,Gimon}. We take the
complex coordinates on $T^2$ to be $z=u+\tau v$ hence we can
define the holomorphic and antiholomorphic one forms that are
elements of the cohomology classes $H^{(1,0)}(T^2,{\mathbb C})$
and $H^{(0,1)}(T^2,{\mathbb C})$ respectively as
 \beq {\Omega}_0=dz=du+\tau dv,\qquad
\overline{\Omega}_0=d\overline{z}=du+\overline{\tau}dv.
\label{omega0} \eeq As it is well-known the Teichm\"uller space of
$T^2$ parametrized by $\tau$ of Eq.(\ref{fura}) has a K\"ahler
metric
$g_{\tau\overline{\tau}}={\partial}_{\tau}{\partial}_{\overline{\tau}}K$
coming from the K\"ahler potential
 \beq
K=-\log(2y).\label{pot}\eeq
 Notice that adopting the convention
$\int_{T^2}du\wedge dv=1$ we have the relation \beq
ie^{-K}=\int_{T^2}\Omega_0\wedge
\overline{\Omega}_0.\label{Kpot}\eeq

Our choice for the volume form on $T^2$ is
\beq\omega=id\overline{z}\wedge dz.\eeq Now the Hodge star is
defined by the formula
$(\varphi,\varphi)\omega=\varphi\wedge\ast\overline{\varphi}$,
hence for $\varphi=\Omega_0=dz$ and its conjugate,
$(\varphi,\varphi)=1$, we get \beq \ast dz=idz,\qquad \ast
d\overline{z}=-id\overline{z}.\label{hodge}\eeq

Let us now define the one-form $\Omega$ as \beq \Omega\equiv
e^{K/2}\Omega_0.\label{nonhol}\eeq\noindent Due to the relations
\beq
(\overline{\tau}-\tau)\left({\partial}_{\tau}+{\partial}_{\tau}K\right)dz=d\overline{z},
\qquad
{\partial}_{\tau}d\overline{z}=0\label{relation}\eeq\noindent 
the flat K\"ahler covariant derivative
defined as
\beq
D_{\hat{\tau}}\Omega\equiv(\overline{\tau}-\tau)D_{\tau}\Omega\equiv
(\overline{\tau}-\tau)\left({\partial}_{\tau}+\frac{1}{2}{\partial}_{\tau}K     \right)
\Omega,\label{covder}\eeq\noindent
\beq
 D_{\hat{\tau}}\overline{\Omega}\equiv
 (\overline{\tau}-\tau)\left({\partial}_{\tau}-\frac{1}{2}{\partial}_{\tau}K
\right) 
 \overline{\Omega}\label{covder2}\eeq\noindent
is acting  as
\beq D_{\hat{\tau}}\Omega=\overline{\Omega},\qquad
D_{\hat{\tau}}\overline{\Omega}=0.\label{actionkov}\eeq\noindent

In order to reinterpret one-forms on $T^2$ as qubits we use the
hermitian inner product \beq \langle \xi\vert\eta\rangle\equiv
\int_{T^2}\xi\wedge\ast\overline{\eta}.\label{inner}\eeq Now one
can show that the correspondence \beq i\Omega\leftrightarrow \vert
0\rangle\qquad i\overline{\Omega}\leftrightarrow \vert
1\rangle\label{corr}\eeq\noindent gives rise to a mapping of basis
states of one-forms to basis states for qubits. By an abuse of
notation we use the same $\langle\vert\rangle$ notation for the
Hermitian inner product on the Hilbert space of qubits i.e. ${\cal
H}\simeq{\mathbb C}^2$ too. Now we have the usual properties
$\langle 0\vert 0\rangle =\langle 1\vert 1\rangle =1$ and $\langle
0\vert 1\rangle =\overline{\langle 1\vert 0\rangle}=0$. By virtue
of this mapping one can reinterpret Eq. (\ref{actionkov}) as \beq
\sigma_+\vert 0\rangle =\vert 1\rangle,\qquad \sigma_{+}\vert
1\rangle =0,\label{projerror}\eeq\noindent i.e. the flat covariant
derivatives act as {\it projective bit flip errors} on the basis
states. Similarly the action of the adjoint of the flat covariant
derivative $D_{\hat{\overline{\tau}}}$ can be reinterpreted as
\beq \sigma_-\vert 0\rangle =0,\qquad \sigma_{-}\vert 1\rangle
=\vert 0\rangle.\label{projerror2}\eeq\noindent Notice also that
our association as given by Eq.(\ref{corr}) represents the
diagonality of the Hodge star operation i.e. $\ast\Omega=i\Omega$,
$\ast\overline{\Omega}=-i\overline{\Omega}$ in the form \beq
\ast\vert 0\rangle =-\vert 0\rangle,\qquad \ast\vert 1\rangle
=+\vert 1\rangle,\label{hodgequbit}\eeq\noindent i.e. the action
of $\ast$ is represented by the sign flip operator $-\sigma_3$.

Now in the context of superstring compactifications the cohomology
classes are {\it real}. By virtue of Poncar\'e duality these
classes are answering the real homology cycles representing brane
configurations wrapped on (for example supersymmetric) cycles. In
the qubit picture this means that our qubits have to satisfy extra
{\it reality conditions}. Moreover, in Calabi-Yau
compactifications self-duality of the usual five-form in the type
IIB duality frame gives a distinguished role to basis states that
diagonalize the Hodge star operator on the Calabi-Yau space. In
this context our torus model should be related to the illustrative
example of Suzuki\cite{Suzuki} where a self-dual three-form was
considered in a compactification model to four space-time
dimensions of the form $M\times T^2$. Hence owing to the special
status of Hodge diagonal states, in the qubit picture we attach to
our basis states $\vert 0\rangle$ and $\vert 1\rangle$ of
Eq.(\ref{corr}) a special role calling them in the following
states of the {\it computational base}.

Let us now write the real cohomology class $\Gamma\in
H^1(T^2,{\mathbb R})$ in the form \beq
\Gamma=p\alpha-q\beta,\qquad \alpha=du,\quad
\beta=dv.\label{realcoh}\eeq\noindent Using the expression
$\Omega_0=\alpha+\tau\beta$ and its conjugate one can express this
in the Hodge diagonal basis as follows \beq
\Gamma=-e^{K/2}(p\overline{\tau}+q) i\Omega+e^{K/2}(p\tau +
q)i\overline{\Omega}.\label{hodgediagcoh}\eeq\noindent According
to our correspondence between one-forms and qubits we can
represent this as a state in the computational base satisfying an
extra reality condition \beq \vert\Gamma\rangle =\Gamma_0\vert
0\rangle +\Gamma_1\vert 1\rangle,\qquad
\Gamma_1=-\overline{\Gamma}_0=e^{K/2}(p\tau+q).\label{qubitform}\eeq\noindent
Notice that although the state itself is not, but both the
amplitudes $\Gamma_{0,1}$ and the (computational) basis vectors
$\vert 0\rangle$ and $\vert 1\rangle $ display an implicit
dependence on the modulus $\tau$. We note also that after imposing
the usual Dirac-Zwanziger quantization condition on $p$ and $q$
$\Gamma$ should rather be interpreted as an element of
$H^1(T^2,{\mathbb Z})$.

Notice also that the state $\vert\Gamma\rangle$ is {\it
unnormalized} with norm squared satisfying \beq
\vert\vert\Gamma\vert\vert ^2=
\langle\Gamma\vert\Gamma\rangle=2e^K\vert
p\tau+q\vert^2=\frac{1}{y}\vert p\tau
+q\vert^2.\label{BHpot1}\eeq\noindent In Quantum Information this
is not a problem since the protocols demanded by quantum
manipulations are not always represented by unitary operators
preserving the norm. In the theory of quantum entanglement one can
consider for instance manipulations converting a state to another
one and vice versa with a probability less than one\cite{Dur}. For
a single qubit these manipulations are represented by the
invertible operations. The nontrivial content of such
manipulations is encapsulated by the group $SL(2,{\mathbb C})$.
For transformations also respecting some additional structure
(e.g. our reality condition) the allowed set of manipulations will
be comprising a subgroup of this group. For our state
$\vert\Gamma\rangle$ it is easy to check that the set of
transformations ${\cal A}$ of the form $\vert\Gamma\rangle\mapsto
{\cal A}\vert \Gamma\rangle$ respecting the reality condition is
comprising the subgroup $SU(1,1)$ of $SL(2,{\mathbb C})$.

Notice also that in matrix representation the state
$\vert\Gamma\rangle$ can be given the form \beq
\begin{pmatrix}\Gamma_0\\\Gamma_1\end{pmatrix}=\frac{1}{\sqrt{2y}}\begin{pmatrix}
\overline{\tau}&-1\\-\tau&1\end{pmatrix}\begin{pmatrix}-p\\q\end{pmatrix}
=\frac{1}{\sqrt{2}}\begin{pmatrix}i&-1\\i&1\end{pmatrix}
\frac{1}{\sqrt{y}}\begin{pmatrix}y&0\\-x&1\end{pmatrix}\begin{pmatrix}-p\\q\end{pmatrix}.
\label{matrixok} \eeq\noindent On the right hand side the first
matrix is unitary, and the second is an element of $SL(2,{\mathbb
R})$. Using this unitary matrix one can switch to another basis
different from our computational one. In this new basis the
subgroup of admissible transformations is $SL(2,{\mathbb R})$. We
also remark that the norm squared $\vert\vert\Gamma\vert\vert ^2$
is a unitary invariant and a symplectic i.e. $SL(2,{\mathbb R})$
one at the same time. The latter invariance means that under the
usual set of combined transformations \beq
\tau\mapsto\frac{a\tau+b}{c\tau+d},\qquad
\begin{pmatrix}-p\\q\end{pmatrix}\mapsto
\begin{pmatrix}d&c\\b&a\end{pmatrix}
\begin{pmatrix}-p\\q\end{pmatrix},\quad ad-bc=1\label{szokotttrafo}\eeq\noindent
the norm squared remains invariant.

Note, that the matrix form of Eq.(\ref{matrixok}) leaves obscure
the fact that the corresponding basis vectors $\vert 0\rangle$ and
$\vert 1\rangle$ are depending on the coordinates of the torus and
the modulus $\tau$. More precisely the set $\{\vert 0\rangle,\vert
1\rangle\}$ refers to {\it families} of basis vectors parametrized
by $\tau$. (The variables $u$ and $v$ on the other hand are
associated with the Hilbert space structure on $T^2$ with inner
product defined by Eq.(\ref{inner}).) Since the possible notation
$\vert 0,1(\tau)\rangle$ , $\Gamma_{0,1}(\tau,p,q)$ displaying
all the implicit structures in Eq.(\ref{qubitform}) is awkward we
leave the symbols $\tau$ and tacitly assume that the
computational basis has an implicit dependence on 
$\tau$. With these conventions our state now has the deceptively
simple appearance

\beq \vert\Gamma\rangle={\cal
S}\vert\gamma\rangle=US\vert\gamma\rangle,\qquad
\vert\gamma\rangle=-p\vert 0\rangle +q\vert
1\rangle\label{simple}\eeq\noindent where the operators ${\cal
S},S,U$ are the ones with matrix representatives easily identified
after looking at Eq.(\ref{matrixok}).

\section{STU model and
three-qubits from $H^3(T^6,{\mathbb C})$}
\subsection{Three qubit
systems}
 The STU model is an $N=2$ supergravity
model\cite{Behrndt} coupled to three vector multiplets interacting
via scalars belonging to the special K\"ahler manifold $[SL(2,
{\mathbb R})/SO(2)]^{\times 3}$. There are many ways embedding
this model to string/M-theory. Here following Borsten
et.al.\cite{Borsten} we use an embedding to type IIB string theory
compactified on the six torus $T^6$, with a three-qubit
interpretation. As was emphasized in that paper\cite{Borsten} the
number of qubits is three because we have now three copies of
$T^2$s corresponding to the six extra dimensions in string theory.
Due to the presence of three qubits here the new phenomenon of
(quantum) entanglement appears, and wrapped $D3$ brane
configurations can effectively be described by such entangled
tripartite states. The aim of the present subsection is to clarify
what do we mean by states in this context, an issue left obscure
in the paper of Borsten et.al.\cite{Borsten}.

In order to do this we just have to generalize our single-qubit
considerations related to $T^2$ known from the previous
subsection, to three-qubits now related to $T^6=T^2\times
T^2\times T^2$. We introduce the coordinates \beq
z^a=u^a+\tau^av^a,\qquad \tau^a=x^a-iy^a\qquad y^a>0,\qquad
a=1,2,3 \label{koordinatak}\eeq\noindent and the holomorphic
three-form \beq \Omega_0=dz^1\wedge dz^2\wedge
dz^3.\label{holthree}\eeq\noindent We have as usual \beq
\int_{T^6}\Omega_0\wedge\overline{\Omega}_0=i(8y^1y^2y^3)=ie^{-K},\label{holnorm}\eeq\noindent
where $K$ is the K\"ahler potential giving rise to the metric
$g_{a\overline{b}}={\partial}_a{\partial}_{\overline{b}}K$ on the
special K\"ahler manifold $[SL(2, {\mathbb R})/SO(2)]^{\times 3}$.
Let us again introduce $\Omega$ as in Eq.(\ref{nonhol}), and
define flat covariant derivatives $D_{\hat{a}}$ acting on $\Omega$
as \beq D_{\hat{a}}\Omega=(\overline{\tau}^a-\tau^a)D_a\Omega=
(\overline{\tau}^a-\tau^a)
\left({\partial}_{a}+\frac{1}{2}{\partial}_aK\right)\Omega,
\label{covder3}\eeq\noindent where
${\partial}_a={\partial}/{\partial{\tau}^a}$. Then one has

\beq \Omega=e^{K/2}dz^1\wedge dz^2\wedge dz^3,\qquad
\overline{\Omega}=e^{K/2}d\overline{z}^1\wedge
d\overline{z}^2\wedge d\overline{z}^3,\label{0cov}\eeq\noindent
\beq D_{\hat{1}}\Omega=e^{K/2}d\overline{z}^1\wedge dz^2\wedge
dz^3,\qquad
\overline{D}_{\hat{\overline{1}}}\overline{\Omega}=e^{K/2}dz^1\wedge
d\overline{z}^2\wedge d\overline{z}^3,\label{1cov}\eeq\noindent
\beq D_{\hat{2}}\Omega=e^{K/2}dz^1\wedge d\overline{z}^2\wedge
dz^3,\qquad
\overline{D}_{\hat{\overline{2}}}\overline{\Omega}=e^{K/2}d\overline{z}^1\wedge
dz^2\wedge d\overline{z}^3,\label{2cov}\eeq\noindent \beq
D_{\hat{3}}\Omega=e^{K/2}dz^1\wedge dz^2\wedge
d\overline{z}^3,\qquad
\overline{D}_{\hat{\overline{3}}}\overline{\Omega}=e^{K/2}d\overline{z}^1\wedge
d\overline{z}^2\wedge dz^3.\label{3cov}\eeq\noindent

Notice that we have the identities \beq
\int_{T^6}\Omega\wedge\overline{\Omega}=i,\qquad
\int_{T^6}D_{\hat{a}}\Omega\wedge
\overline{D}_{\hat{\overline{b}}}\overline{\Omega}=-i
{\delta}_{\hat{a}{\hat{\overline{{b}}}}}.\label{covnorm}
\eeq\noindent

Let us revisit\cite{Denef} the action of the Hodge star on our
basis of three-forms as given by Eq.(\ref{0cov})-(\ref{3cov}). For
a form of $(p,q)$ type the action of the Hodge star is defined as
\beq
(\varphi,\varphi)\frac{\omega^n}{n!}=\varphi\wedge\ast\overline{\varphi}
\label{hodgedef}\eeq\noindent where for our $T^6$ in accord with
our conventions \beq \omega=i(d\overline{z}^1\wedge dz^1+
d\overline{z}^2\wedge dz^2+d\overline{z}^3\wedge
dz^3)\label{omegat6}\eeq\noindent moreover, we have \beq
(\varphi,\varphi)\equiv\frac{1}{p!q!}\sum \vert \varphi_{j_1\dots
j_pk_1\dots k_q}\vert^2.\label{finorm}\eeq\noindent For our basis
forms like $\varphi\equiv dz^1\wedge dz^2\wedge dz^3$ e.t.c.
showing up in Eq.(\ref{0cov})-(\ref{3cov}) $(\varphi,\varphi)=1$
hence we get \beq \ast\Omega=i\Omega,\qquad
\ast\overline{\Omega}=-i\overline{\Omega}\label{elsohodge}\eeq\noindent
\beq \ast D_{\hat{a}}\Omega=-iD_{\hat{a}},\qquad
\ast\overline{D}_{\hat{\overline{a}}}\overline{\Omega}=i\overline{D}_{\hat{\overline{a}}}
\overline{\Omega}\label{masodikhodge}\eeq\noindent i.e. our
conventions are differing by a sign from the ones of
Denef\cite{Denef}.

Now we regard the $8$ complex dimensional untwisted primitive
part\cite{Lust} of the $20$ dimensional space $H^3(T^6,{\mathbb
C})\equiv H^{3,0}\oplus H^{2,1}\oplus H^{1,2}\oplus H^{0,3}$
equipped with the Hermitian inner product \beq
\langle\varphi\vert\eta\rangle\equiv\int_{T^6}
\varphi\wedge\ast\overline{\eta}\label{inner3}\eeq\noindent as a
Hilbert space isomorphic to ${\cal H}\equiv ({\mathbb
C}^2)^{\times 3}\simeq {\mathbb C}^8$ of three qubits. In order to
set up the correspondence between the three-forms and the basis
vectors of the three-qubit system we use the {\it negative} of the
basis vectors $\Omega,D_{\hat{1}}\Omega$ etc. multiplied by the
imaginary unit $i$. We opted for using an extra minus sign since
after changing the order of the one-forms we have for example for
$ -iD_{\hat{1}}\Omega=ie^{K/2}dz^3\wedge dz^2\wedge
d\overline{z}^1$ hence we can take its representative basis qubit
state $\vert 001\rangle$ which corresponds to the usual binary
labelling provided we label the qubits from the right to the left.
Due to these conventions we take the basis states of our
computational base to be given by the correspondence \beq
-i\Omega\leftrightarrow \vert 000\rangle,\qquad
-iD_{\hat{1}}\Omega\leftrightarrow \vert 001\rangle, \qquad
-iD_{\hat{2}}\Omega\leftrightarrow \vert 010\rangle,\qquad
-iD_{\hat{3}}\Omega\leftrightarrow \vert
100\rangle\label{qubitcorrespondence}\eeq\noindent \beq
-i\overline{\Omega}\leftrightarrow \vert 111\rangle,\qquad
-i\overline{D}_{\hat{\overline{1}}}\Omega\leftrightarrow \vert
110\rangle, \qquad
-i\overline{D}_{\hat{\overline{2}}}\Omega\leftrightarrow \vert
101\rangle,\qquad
-i\overline{D}_{\hat{\overline{3}}}\Omega\leftrightarrow \vert
011\rangle.\label{qubitcorrespondence2}\eeq\noindent Now the
locations of the $1$s correspond to the slots where complex
conjugation is effected. One can check that the states above form
a basis with respect to the inner product of Eq.(\ref{inner3})
with the usual set of properties on the three-qubit side. A
further check shows that the action of the flat covariant
derivatives $D_{\hat{a}}, j=1,2,3$ corresponds to the action of
the projective bit flips of the form $I\otimes
I\otimes\sigma_+,\quad I\otimes\sigma_+\otimes I$ and
$\sigma_+\otimes I\otimes I$, where $I$ is the $2\times 2$
identity matrix. For the conjugate flat covariant derivatives
$\sigma_+$ has to be replaced by ${\sigma}_-$. Moreover, the
diagonal action of the Hodge star in the computational base is
represented by the corresponding action of the {\it negative} of
the parity check operator $\sigma_3\otimes\sigma_3\otimes
\sigma_3$.

Now for a three-form representing the cohomology class of a
wrapped $D3$ brane configuration we take  \beq
\Gamma=p^I\alpha_I-q_I\beta^I\in H^3(T^6,{\mathbb
Z}),\label{threebrane}\eeq\noindent with summation on $I=0,1,2,3$
and \beq \alpha_0=du^1\wedge du^2\wedge du^3,\qquad
\beta^0=-dv^1\wedge dv^2\wedge dv^3\label{nullasok}\eeq\noindent
\beq \alpha_1=dv^1\wedge du^2\wedge du^3,\qquad \beta^1=du^1\wedge
dv^2\wedge dv^3\label{egyesek}\eeq\noindent with the remaining
ones obtained via cyclic permutation. With the choice of
orientation $\int_{T^6}(du^1\wedge dv^1)\wedge (du^2\wedge
dv^2)\wedge(du^3\wedge dv^3)=1$ we have
$\int_{T^6}{\alpha}_I\wedge {\beta}^J={\delta}_I^J$.

It is well-known\cite{Denef} that in the Hodge diagonal basis we
can express this as \beq \Gamma=iZ(\Gamma)\overline{\Omega}
-ig^{j\overline{k}}D_jZ(\Gamma)
\overline{D}_{\overline{k}}\overline{\Omega}+{\rm c.c.}=
iZ(\Gamma)\overline{\Omega}
-i\delta^{\hat{j}\hat{\overline{k}}}D_{\hat{j}}Z(\Gamma)
\overline{D}_{\hat{\overline{k}}}\overline{\Omega}+{\rm
c.c.}\label{expand}\eeq\noindent Here
$Z(\Gamma)=\int_{T^6}\Gamma\wedge\Omega$ is the central charge.
For the STU model the explicit form of $Z(\Gamma)$
is\cite{Behrndt} \beq
Z(\Gamma)=e^{K/2}W(\tau^3,\tau^2,\tau^1)\label{centszuper}\eeq\noindent
where \beq W(\tau^3,\tau^2,\tau^1)=q_0+q_1\tau^1
+q_2\tau^2+q_3\tau^3+p^1\tau^2\tau^3+p^2\tau^1\tau^3+p^3\tau^1\tau^2-p^0\tau^1\tau^2\tau^3.
\label{centralcharge}\eeq\noindent

Now using our basic correspondence between three-forms and
three-qubit states of Eq.
(\ref{qubitcorrespondence})-(\ref{qubitcorrespondence2}) we can
write $\Gamma \leftrightarrow \vert\Gamma\rangle$ where \beq
\vert\Gamma\rangle=\Gamma_{000}\vert 000\rangle +\Gamma_{001}\vert
001\rangle +\dots +{\Gamma}_{110}\vert 110\rangle
+{\Gamma}_{111}\vert 111\rangle,\label{qubitform2}\eeq\noindent
where \beq
{\Gamma}_{111}=-e^{K/2}W(\tau^3,\tau^2,\tau^1)=-\overline{\Gamma}_{000},\label{ampl1}\eeq\noindent
\beq
{\Gamma}_{001}=-e^{K/2}W(\overline{\tau}^3,\overline{\tau}^2,\tau^1)=-\overline{\Gamma}_{110}
\label{ampl2} \eeq\noindent and the remaining amplitudes are given
by cyclic permutation.

Let us now put the $8$ charges $p^I$ and $q_I$ with $I=0,1,2,3$ to
a $2\times 2\times 2 $ array $\gamma_{kji}$ $k,j,i=0,1$ as follows
\beq
\begin{pmatrix}\gamma_{000}&\gamma_{001}&\gamma_{010}&\gamma_{100}\\
\gamma_{111}&\gamma_{110}&\gamma_{101}&\gamma_{011}\end{pmatrix} =
\begin{pmatrix} -p^0&-p^1&-p^2&-p^3\\-q_0&q_1&q_2&q_3\end{pmatrix}.
\label{toltesmatrix}\eeq\noindent Now it can be shown that the
three-qubit state of Eq.(\ref{qubitform2}) can alternatively be
written in the following form \beq \vert\Gamma\rangle={\cal
S}_3\otimes{\cal S}_2\otimes {\cal S}_1\vert
\gamma\rangle\label{statecohomology}\eeq\noindent where
\beq\vert\gamma\rangle=\gamma_{000}\vert
000\rangle+\gamma_{001}\vert 001\rangle+\dots +\gamma_{110}\vert
110\rangle +\gamma_{111}\vert
111\rangle,\label{kisgamma}\eeq\noindent and the matrix
representative of the operator ${\cal S}_3\otimes {\cal
S}_2\otimes {\cal S}_1$ is \beq\frac{1}{\sqrt{8y^3y^2y^1}}
\begin{pmatrix}\overline{\tau}^3&-1\\-\tau^3&1\end{pmatrix}\otimes
\begin{pmatrix}\overline{\tau}^2&-1\\-\tau^2&1\end{pmatrix}\otimes
\begin{pmatrix}\overline{\tau}^1&-1\\-\tau^1&1\end{pmatrix}.
\label{matricak}\eeq\noindent The reader should compare this expression with the
one obtained for the single qubit case as shown by
Eqs.(\ref{matrixok}) and (\ref{simple}).

A state similar to $\vert\Gamma\rangle$ of
Eq.(\ref{statecohomology}) has already appeared in our recent
papers\cite{Levay1}. It is important to realize however, the basic
difference between $\vert\Gamma\rangle$ and that state. The state
of Ref.\cite{Levay1} is a {\it charge} and {\it moduli} dependent
state connected to the $4$ dimensional setting of the STU model.
Moreover, in that setting the basis states $\vert kji\rangle$ had
no obvious physical meaning. They merely served as basis vectors
providing a suitable frame for a three-qubit reformulation.

Now $\vert\Gamma\rangle$ is a state which is depending on the {\it
charges} the {\it moduli} and the {\it coordinates of the extra
dimensions}, hence this state is connected to a $10$ dimensional
setting of the STU model in the type IIB duality frame. Now the
basis vectors $\vert kji\rangle$ have an obvious physical meaning:
they are the Hodge diagonal complex basis vectors of the untwisted primitive part of the
third cohomology group of the extra dimensions i.e. of $H^3(T^6,{\mathbb
C})$. They are also basis vectors of a genuine Hilbert space
equipped with a natural Hermitian inner product of
Eq.(\ref{inner3}), isomorphic to the usual one of three-qubits.
The state $\vert\Gamma\rangle$ has the meaning as the Poincar\'e dual of the
homology cycle representing wrapped $D3$ brane configurations.
$\vert\Gamma\rangle$ can be represented in two different forms:
namely as in Eq.(\ref{qubitform2}) (expansion in a Hodge-diagonal
moduli dependent complex base), or in an equivalent way based on
the qubit version of Eq.(\ref{threebrane}) (Hodge-non-diagonal but
moduli independent real base).

In closing this section we present the analogue of
Eq.(\ref{BHpot1}) i.e. the norm of $\vert\Gamma\rangle$ \beq
\vert\vert\Gamma\vert\vert^2=2e^K(\vert
W(\tau^3,\tau^3,\tau^1)\vert^2+\vert
W(\overline{\tau}^3,\tau^2,\tau^1)\vert^2+
W(\tau^3,\overline{\tau}^2,\tau^1)\vert^2+\vert
W(\tau^3,\tau^2,\overline{\tau}^1)\vert^2).\label{blackholepotential}\eeq\noindent This
expression is just $2$ times  $V_{BH}$, the well-known black hole
potential\cite{Ferrarastu}. For a three-qubit based reformulation
of $V_{BH}$ see Ref.\cite{Levay1} Now its new interpretation as
half the norm of a three-qubit state involves integration with
respect to the coordinates of the extra dimensions (see
Eq.(\ref{inner3}). It is obvious by construction that $V_{BH}$ is
a unitary and symplectic invariant ($SL(2,{\mathbb R}^{\times
3}\subset Sp(8,{\mathbb R})$) at the same time. In order to see
this one just has to recall our considerations for the single
qubit case encapsulated in Eqs. (\ref{matrixok})
-(\ref{szokotttrafo}).

\subsection{BPS attractors} As a first application showing the
usefulness of rephrasing well-known results concerning the STU
model in a three-qubit language let us consider the case of the
BPS attractors\cite{Behrndt,Ferrarastu}. In this case the BPS
conditions read as (which is also a requirement of unbroken
supersymmetry) \beq D_aZ=0.\label{BPS}\eeq\noindent Let us write
the superpotential in the three-qubit form as
 \beq
W(\tau^3,\tau^2,\tau^1)={\Gamma}_{kji}c^kb^ja^i=\Gamma_{kji}\varepsilon^{ii^{\prime}}
\varepsilon^{jj^{\prime}}\varepsilon^{kk^{\prime}}
c_{k^{\prime}}b_{j^{\prime}}a_{i^{\prime}} \label{wmaskepp}\eeq
\noindent where summation over
$i^{\prime},j^{\prime},k^{\prime}=0,1$ is understood and
$\varepsilon^{01}=-\varepsilon^{10}=1$ are the nonzero components
of the the usual $SL(2)$ invariant $2\times 2$ matrix, and \beq
a_i\leftrightarrow
\begin{pmatrix}1\\\tau^1\end{pmatrix},\qquad
b_j\leftrightarrow
\begin{pmatrix}1\\\tau^2\end{pmatrix}\qquad c_k\leftrightarrow
\begin{pmatrix}1\\\tau^3\end{pmatrix}.
\label{vektorok}\eeq\noindent Then the BPS attractors are
characterized by the equations \beq
W(\overline{\tau}^3,\tau^2,\tau^1)=0,\qquad
W(\tau^3,\overline{\tau^2},\tau^1)=0,\qquad
W(\tau^3,\tau^2,\overline{\tau^1})=0\label{BPSsuper}\eeq\noindent
and their complex conjugates. According to
Eqs.(\ref{ampl1})-(\ref{ampl2}) in our three-qubit language this
corresponds to \beq
\Gamma_{001}=\Gamma_{010}=\Gamma_{100}=\Gamma_{110}=\Gamma_{101}=\Gamma_{011}=0.
\label{30feltetel}\eeq\noindent This means that at the black hole
horizon after moduli stabilization only the ${\Gamma}_{000}$ and
${\Gamma}_{111}$ amplitudes of our state $\vert\Gamma\rangle$
survives. Hence the unfolding of the attractor flow towards its
fixed point can be reinterpreted as a distillation
procedure\cite{Levay1} of a GHZ state of the form \beq
\vert\Gamma\rangle_{fix}\equiv \Gamma_{000}\vert
000\rangle_{fix}+\Gamma_{111}\vert 111\rangle_{fix},\qquad
\Gamma_{111}=-\overline{\Gamma}_{000}=Z(\tau^3_{fix},\tau^2_{fix},\tau^1_{fix};p,q).
\label{GHZ}\eeq\noindent Notice that this known  result in our new
interpretation directly relates the distillation procedure to the
well-known property of supersymmetric attractors in the type IIB
picture namely that in this case only the $H^{3,0}$ and
$H^{0,3}$ parts of the cohomology survive\cite{Moore}.

In order to present the usual solution for the fixed values of the
moduli write Eqs.(\ref{BPSsuper}) in the form \beq
\Gamma_{kji}\overline{c}^kb^ja^i=0,\qquad
\Gamma_{kji}c^k\overline{b}^ja^i=0,\qquad
\Gamma_{kji}c^kb^j\overline{a}^i=0.
\label{maskeppsuper}\eeq\noindent Using the fact that
$\Gamma_{kji}$ is real these equations taken together with their
complex conjugates are equivalent to the vanishing of the $2\times
2$ determinants\cite{Behrndt,Levay1} \beq {\rm
Det}\left(\Gamma_{kji}c^k\right)=0,\qquad {\rm Det}
\left(\Gamma_{kji}b^j\right)=0,\qquad{\rm
Det}\left(\Gamma_{kji}a^i\right)=0,\label{megoldas}\eeq\noindent
provided the imaginary parts of the moduli are non vanishing. (A
property clearly should hold due to physical reasons.) The above
equations result in three quadratic equations, keeping only the
solutions providing $y^1,y^2$ and $y^3$ positive yield the
stabilized  values for the moduli\cite{Behrndt,Levay1,Ferrarastu}
\beq\tau^a_{fix}=\frac{(\gamma_0\cdot\gamma_1)^a+i\sqrt{-D}}
{(\gamma_0\cdot\gamma_0)^a},\qquad
a=1,2,3.\label{ittaveg}\eeq\noindent Here for example
\beq(\gamma_0\cdot\gamma_1)^1\equiv
\gamma_{kj0}\varepsilon^{kk^{\prime}}\varepsilon^{jj^{\prime}}
\gamma_{k^{\prime}j^{\prime}1},\label{szorzat}\eeq\noindent and
\beq D=(\gamma_0\cdot\gamma_1)^2-(\gamma_0\cdot\gamma_0)
(\gamma_1\cdot\gamma_1)\label{Cayley}\eeq\noindent is Cayley's
hyperdeterminant\cite{Cayley,Duff}. In order to have such
solutions $-D$ should be positive and $(\gamma_0\cdot\gamma_0)$
should both be negative\cite{Ferrarastu}.

Using the stabilized values ${\tau}^a_{fix},\quad a=1,2,3$ in
$e^{K/2}W(\tau^3,\tau^2,\tau^1)$ we get the well-known
result\cite{Behrndt,Duff,Levay1} \beq \vert Z\vert ^2= e^{K}\vert
W(\tau^3_{fix},\tau^2_{fix},\tau^1_{fix})\vert^2
=\sqrt{-D(\vert\gamma\rangle)}=\sqrt{
(\gamma_0\cdot\gamma_0)(\gamma_1\cdot\gamma_1)-(\gamma_0\cdot\gamma_1)^2}\label{centralcayley}
\eeq\noindent where due to the triality symmetry of Cayley's
hyperdeterminant products like $(\gamma_0\cdot\gamma_1)$ can be
calculated by using any of the qubits playing a special role. This
quantity is showing up in the macroscopic Bekenstein-Hawking
entropy of the extremal, static, spherical symmetric BPS black
hole solution of the
STU-model\cite{Behrndt,Ferrarastu,Duff,Levay1} \beq
S_{BH}=\pi\sqrt{-D(\vert\gamma\rangle)}.\label{entropy}\eeq\noindent
Note that the quantity ${\tau_3}=4\vert
D(\vert\gamma\rangle)\vert$ is a genuine entanglement measure of
the state $\vert\gamma\rangle$ in the theory of three-qubit
entanglement\cite{Kundu}. For BPS black holes we have
$\tau_3=-4D$.

The final form of our three-qubit state on the horizon of the
black hole is \beq
\vert\Gamma\rangle_{fix}=(-D)^{1/4}\left(e^{i\alpha}\vert
000\rangle_{fix} -e^{-i\alpha}\vert
111\rangle_{fix}\right),\label{finalstate}\eeq\noindent where\beq
\tan\alpha=\sqrt{-D}\frac{p^0}{2p^1p^2p^3+p^0(p^0q_0+p^1q_1+p^2q_2+p^3q_3
) }.\label{fazis}\eeq\noindent As we see this unnormalized state
is of generalized GHZ form\cite{GHZ}, where the relative phase is
given by the phase of the central charge. Hence the attractor
mechanism can be regarded as a distillation procedure of a GHZ
state on the black hole horizon\cite{Levay1}. However, as a new
result here one should also see that according to our basic
correspondence between cohomology classes and qubits the vectors
$\vert 000\rangle_{fix}$ and $\vert 111\rangle_{fix}$ now
correspond to the covariantly holomorphic and antiholomorphic
three-forms $-i\Omega_{fix}$ and $-i\overline{\Omega}_{fix}$
respectively.

\section{A IIB $(T^2)^3/({\mathbb Z}_2\times{\mathbb Z}_2)$ model
for flux compactification}
\subsection{Four qubit systems}
In this section we show yet another application of the qubit
picture connected to flux compactification. In order to do this
first we have to connect our considerations to {\it four} qubit
systems.

First we combine the type IIB NS and RR three-forms $H_3$ and
$F_3$ into a new three-form $G_3$ which has also a dependence on a
special type of new moduli $\tau=a+ie^{-\Phi}$ i.e. the axion
dilaton field. The usual expression of $G_3$ is \beq G_3=F_3-\tau
H_3.\label{G3}\eeq\noindent

Now we embed our type $IIB$ model based on the space
$Y=(T^2)^3/({\mathbb Z}_2\times{\mathbb Z}_2)$ (and restricting merely to the untwisted sector) into F-theory on an
elliptically fibered CY four-fold. It is convenient to introduce a
four-form $G_4$ via making use of an {\it extra} torus $T^2$ as
follows. Define \beq G_4=dv\wedge F_3+du\wedge
H_3=\frac{1}{\tau-\overline{\tau}}(G_3\wedge
d\overline{z}-\overline{G}_3\wedge dz),\label{G4}\eeq\noindent
where  $du$ and $dv$ are the coordinates of the new torus $T^2$
with $dz= du+\tau dv=\alpha+\tau\beta$, with K\"ahler potential
$K_1=-\log i(\overline{\tau}-\tau)$. We still have the property
$\int_{T^2}\alpha\wedge\beta =1$. Notice that $G_4$ is invariant
under the $SL(2,{\mathbb R})$ duality symmetry of the $IIB$
theory. This means that under the transformations \beq
\tau\mapsto\frac{a\tau+b}{c\tau+d},\qquad
G_3\mapsto\frac{1}{c\tau+d}G_3,\qquad
dz\mapsto\frac{1}{c\tau+d}dz,\label{trafok2}\eeq\noindent
originating from the set of transformations \beq
\begin{pmatrix}H\\F\end{pmatrix}\mapsto\begin{pmatrix}d&c\\b&a\end{pmatrix}
\begin{pmatrix}H\\F\end{pmatrix},\qquad \begin{pmatrix}\alpha\\\beta\end{pmatrix}\mapsto
\begin{pmatrix}a&-b\\-c&d\end{pmatrix}
\begin{pmatrix}\alpha\\\beta\end{pmatrix}.
\label{masiktr2}\eeq\noindent $G_4$ is left invariant.

Using the form of the K\"ahler potential
$K_1=-\log(\overline{\tau}-\tau)$ we notice that the $G_4$ can be
reinterpreted as a four-qubit state. Indeed $G_4$ can be regarded
as the sum of two components that can be put into a two component
vector as \beq ie^{K_1/2}\begin{pmatrix}
-\overline{\tau}&1\\\tau&-1\end{pmatrix}\begin{pmatrix}H_3\wedge
d\overline{z}\\F_3\wedge
dz\end{pmatrix}.\label{matrixg4}\eeq\noindent Let us now use the
expressions in the Hodge diagonal base

\beq H_3=P^I\alpha_I-Q_I\beta^I= iZ(H)\overline{\Omega} -i\delta
^{\hat{j}\hat{\overline{k}}}D_{\hat{j}}Z(H)
\hat{D}_{\hat{\overline{k}}}\overline{\Omega}+{\rm
c.c.}\label{expandH}\eeq\noindent

\beq F_3=p^I\alpha_I-q_I\beta^I= iZ(F)\overline{\Omega}
-i\delta^{\hat{j}\hat{\overline{k}}}D_{\hat{j}}Z(F)
\overline{D}_{\hat{\overline{k}}}\overline{\Omega}+{\rm
c.c.}\label{expandF}\eeq\noindent

According to Section III. we know that to these expressions one
can associate a pair of three-qubit states as

\beq \vert H\rangle=H_{000}\vert 000\rangle +H_{001}\vert
001\rangle +\dots +{H}_{110}\vert 110\rangle +{H}_{111}\vert
111\rangle,\label{qubitformH}\eeq\noindent

\beq \vert F\rangle=F_{000}\vert 000\rangle +F_{001}\vert
001\rangle +\dots +{F}_{110}\vert 110\rangle +{F}_{111}\vert
111\rangle.\label{qubitformF}\eeq\noindent

In order to fit these states into a four qubit one we need some
minor adjustments. According to Eq.(\ref{fura}) we have chosen
moduli to have negative imaginary parts, however $\tau$ has
positive imaginary part. Moreover, the complex differential
associated to $dz=\alpha+\tau\beta$ was featuring $\tau$. We can
regard all moduli on the same footing by defining a fourth moduli
and the complex coordinate of the associated torus as \beq
\tau^4\equiv\overline{\tau},\qquad dz^4\equiv
d\overline{z}.\label{atteres}\eeq\noindent

Let us now define the covariantly holomorphic four-form as \beq
\Omega=e^{{\cal K}/2}dz^4\wedge dz^3\wedge dz^2\wedge
dz^1\label{covholfour}\eeq\noindent where the total K\"ahler
potential is ${\cal K}\equiv K_1+K$ with $K$ showing up in
Eq.(\ref{holnorm}). Once again we define flat covariant
derivatives $D_{\hat{A}}$, $A=1,2,3,4$ and the quantities
$D_{\hat{A}}\Omega$. The conjugate quantities will be denoted as
usual by $\overline{\Omega}$ and
$\overline{D}_{\hat{\overline{A}}}\overline{\Omega}$. Then we have
for example \beq D_{\hat{4}}\Omega=e^{{\cal
K}/2}d\overline{z}^4\wedge dz^3\wedge dz^2\wedge
dz^1.\label{cov4hatas}\eeq\noindent

Now one has the expansion\cite{Denefdouglas,Kalloshnew} for $G_4$
as an element of the space of allowed fluxes $H_G^4(T^2\times Y)$
\beq G_4= Z(G)\Omega-\overline{D}^{\hat{A}}\overline{Z}
(G)D_{\hat{A}}\Omega+
\overline{D}^{\hat{4}\hat{I}}\overline{Z}(G)D_{\hat{4}\hat{I}}+c.c\label{douglas}\eeq\noindent

We can reinterpret this expansion as a state $\vert G\rangle$
satisfying the usual reality condition in $({\mathbb C}^2)^{\times
4}$ if we make the correspondence \beq \vert
0000\rangle\leftrightarrow \Omega, \qquad\vert 1111\rangle
\leftrightarrow \overline{\Omega},\label{015}\eeq\noindent
\beq\vert 0001\rangle\leftrightarrow D_{\hat{1}}\Omega,\qquad
\vert 1110\rangle\leftrightarrow
\overline{D}_{\hat{\overline{1}}}\overline{\Omega},\dots{\rm
e.t.c.}\label{114}\eeq\noindent \beq \vert
1001\rangle\leftrightarrow D_{\hat{4}}D_{\hat{1}}\Omega,\qquad
\vert 0110\rangle\leftrightarrow
\overline{D}_{\hat{\overline{4}}}\overline
D_{\hat{\overline{1}}}\overline{\Omega},\dots {\rm e.t.c.}
\label{96}\eeq\noindent Now the expansion in the Hodge diagonal
basis is having the alternative form of a $4$-qubit state \beq
\vert G\rangle =G_{0000}\vert 0000\rangle +G_{0001}\vert
0001\rangle +\dots G_{1110}\vert 1110\rangle +G_{1111}\vert
1111\rangle.\label{4qbitform}\eeq\noindent Recall again that in
the Hodge diagonal basis the operator $\ast$ is acting as the
parity check operator
$\sigma_3\otimes\sigma_3\otimes\sigma_3\otimes\sigma_3$, and the
flat covariant derivatives and their conjugates act as suitable
numbers of $\sigma_+$ or $\sigma_-$ operators inserted in fourfold
tensor products.

Notice that the state $\vert G\rangle$ can be written in the form
\beq \vert G\rangle={\cal S}_4\otimes{\cal S}_3\otimes {\cal
S}_2\otimes {\cal S}_1\vert g\rangle\label{fluxstate}\eeq\noindent
where the matrix representative of the four-fold tensor product of
operators is\beq\frac{1}{\sqrt{16y^4y^3y^2y^1}}
\begin{pmatrix}\overline{\tau}^4&-1\\-\tau^4&1\end{pmatrix}\otimes
\begin{pmatrix}\overline{\tau}^3&-1\\-\tau^3&1\end{pmatrix}\otimes
\begin{pmatrix}\overline{\tau}^2&-1\\-\tau^2&1\end{pmatrix}\otimes
\begin{pmatrix}\overline{\tau}^1&-1\\-\tau^1&1\end{pmatrix},\label{negyesszorzat}\eeq\noindent
and the flux state $\vert g\rangle$ is defined as \beq \vert
g\rangle=\sum_{lkji=0,1}g_{lkji}\vert
lkji\rangle,\label{kisfluxstate}\eeq\noindent with the explicit
form of the amplitudes is given by
 \beq
\begin{pmatrix}g_{0000}&g_{0001}&g_{0010}&g_{0100}\\g_{0111}&g_{0110}&g_{0101}&g_{0011}
\end{pmatrix}
=\begin{pmatrix}-p^0&-p^1&-p^2&-p^3\\-q_0&q_1&q_2&q_3\end{pmatrix}
\label{flux1charge}\eeq\noindent

\beq
\begin{pmatrix}g_{1000}&g_{1001}&g_{1010}&g_{1100}\\g_{1111}&g_{1110}&g_{1101}&g_{1011}
\end{pmatrix}
=\begin{pmatrix}-P^0&-P^1&-P^2&-P^3\\-Q_0&Q_1&Q_2&Q_3\end{pmatrix}.
\label{flux2charge}\eeq\noindent

Here the amplitudes containing the  fluxes $(p^I,q_I)$ and
$(P^I,Q_I)$ are just the ones appearing in
Eqs.(\ref{expandH})-(\ref{expandF}).

\subsection{Flux attractors}

In this subsection as an illustration we study an
example\cite{OConnell} of the attractor equation of flux
compactification on the orbifold $T^6/{\mathbb Z}_2\times{\mathbb
Z}_2$. In this case the flux attractor equations are just a
rephrasing of the imaginary self duality condition\cite{Dasgupta}
(ISD) $\ast_6 G=iG$ for the complex flux form of Eq.(\ref{G3}).
This condition arising from the $10D$ equations of motion imply
that the complex structure of the Calabi-Yau space is fixed in a
way such that $G_3$ has only $(0,3)$ and $(2,1)$ components. It is
also known that the ISD condition is equivalent to the ones of
$D_i{\cal W}=D_{\tau}{\cal W}=0$ where ${\cal
W}=\int_{CY}G_3\wedge \Omega_3$ is the GVW superpotential. Here
$D_i$ and $D_{\tau}$ are the covariant derivatives featuring the
complex structure moduli and the complex axio-dilaton. As
discussed in the previous section in the four-qubit formalism
based on the four-form $G_4$ and its associated state $\vert
G\rangle$ these conditions boil down to the ones \beq
G_{0001}=G_{0010}=G_{0100}=G_{1000}=G_{1110}=G_{1101}=G_{1011}=G_{0111}=0.\label{4feltetel}
\eeq\noindent \ Recall that \beq G_{0000}=Z(G)=\int_{Y\times
T^2}G_4\wedge
\Omega=e^{K/2}W(\tau^4,\tau^3,\tau^2,\tau^1),\label{central4}\eeq\noindent
where \begin{eqnarray}
W&=&-q_0-q_1\tau^1-q_2\tau^2-q_3\tau^3+Q_0\tau^4-p^1\tau^2\tau^3-p^2\tau^1\tau^3-p^3\tau^1\tau^2
+Q_1\tau^1\tau^4+Q_2\tau^2\tau^4\nonumber\\&+&
Q_3\tau^3\tau^4+p^0\tau^1\tau^2\tau^3+P^1\tau^2\tau^3\tau^4+P^2\tau^1\tau^3\tau^4
+P^3\tau^1\tau^2\tau^4-P^0\tau^1\tau^2\tau^3\tau^4.\label{Wreszletes}
\end{eqnarray}\noindent
Moreover, according to our interpretation of the action of the
flat covariant derivatives as projective bit flip errors
amplitudes like $G_{0001}$ are just obtained from the expression
of $G_{0000}$ by replacing the corresponding moduli by its complex
conjugate in the relevant slot hence for example we have
$G_{0001}=e^{K/2}W(\tau^4,\tau^3,\tau^2,\overline{\tau}^1)$. Hence
in this four-qubit reinterpretation the flux attractor equations
again correspond to some distillation procedure of our state
$\vert G\rangle$ where from the $16$ amplitudes due to the
vanishing of the ones of Eq.(\ref{4feltetel}) only $8$ ones will
survive.

In order to illustrate this distillation procedure in detail we
invoke the explicit solution found by Larsen and
OConnell\cite{OConnell}. This solution is a one with merely $8$
fluxes, i.e. in the definition of $\vert G\rangle$ one takes \beq
\begin{pmatrix}g_{0000}&g_{0001}&g_{0010}&g_{0100}\\g_{0111}&g_{0110}&g_{0101}&g_{0011}
\end{pmatrix}
=\begin{pmatrix}-p^0&0&0&0\\0&q_1&q_2 &q_3\end{pmatrix}
\label{flux1charge2}\eeq\noindent \beq
\begin{pmatrix}g_{1000}&g_{1001}&g_{1010}&g_{1100}\\g_{1111}&g_{1110}&g_{1101}&g_{1011}
\end{pmatrix}
=\begin{pmatrix}0&-P^1&-P^2&-P^3\\-Q_0&0&0&0\end{pmatrix}.
\label{flux2charge2}\eeq\noindent Using a generating function for
the flux attractor equations in Ref.\cite{OConnell} the authors
have shown that this configuration with $8$ fluxes has a purely
imaginary solution for the four moduli $\tau^a$ of the form \beq
\tau^1=-i\left(-\frac{Q_0P^1q_2q_3}{P^2P^3p^0q_1}\right)^{1/4},\qquad
\tau^2=-i\left(-\frac{Q_0P^2q_1q_3}{P^1P^3p^0q_2}\right)^{1/4},\label{12mod}
 \eeq\noindent
\beq
\tau^3=-i\left(-\frac{Q_0P^3q_1q_2}{P^1P^2p^0q_3}\right)^{1/4},\qquad
\tau^4=-i\left(-\frac{p^0q_1q_2q_3}{Q_0P^1P^2P^3}\right)^{1/4},\label{34mod}
 \eeq\noindent
\beq -{\rm sgn}(Q_0p^0)={\rm sgn}(P^1q_1)={\rm sgn}(P^2q_2)={\rm
sgn}(P^3q_3)=+1. \label{signconve}\eeq\noindent Recall that
$\tau^4=\overline{\tau}$ where $\tau=C_0+ie^{-\phi}$ is the
axio-dilaton. Now $C_0=0$ hence $-{\tau}^4$ gives the stabilized
value of the dilaton.

One can easily check that these stabilized values indeed satisfy
the constraints of Eq.(\ref{4feltetel}). In order to do this just
write\beq W=(p^0\tau^1\tau^2\tau^3+ Q_0\tau^4 )+
(P^1\tau^2\tau^3\tau^4-q_1\tau^1)+(P^2\tau^1\tau^3\tau^4
-q_2\tau^2)+ (P^3\tau^1\tau^2\tau^4
 -q_3\tau^3)
\eeq \noindent and check that the terms in the brackets give zero
when we conjugate an {\it odd} number of moduli in the expression
of $W$. In order to reveal the distillation procedure at work let
us first calculate $\vert G\rangle_{fix}$ using these stabilized
values for the moduli. We introduce the quantities \beq x={\rm
sgn}(q_1)\sqrt{P^1q_1},\quad y={\rm sgn}(q_2)\sqrt{P^2q_2},\quad
z={\rm sgn}(q_3)\sqrt{P^3q_3},\quad t=-{\rm
sgn}(-Q_0)\sqrt{-Q_0p^0}.\label{mennyisegek}\eeq\noindent Then a
calculation shows that for $l,k,j,i\in\{0,1\}$ \beq
\left(G_{lkji}\right)_{fix}=\frac{i}{2}\left[(-1)^lt+(-1)^kz+(-1)^jy+(-1)^ix\right],\qquad
l+k+j+i\equiv 0{\rm mod}2,\label{amplitudok}\eeq\noindent and of
course due to Eq.(\ref{4feltetel}) we have
 \beq \left(G_{lkji}\right)_{fix}=0,\qquad l+k+j+i=1{\rm
mod}2.\label{amplitudok2}\eeq\noindent

A quantity of physical importance which is related to one of these
nonzero amplitudes is the complex gravitino mass $M_{3/2}$. This
quantity is depending on the fluxes and the moduli. Its explicit
form at the attractor point is given by the formula \beq
(M_{3/2})_{fix}^2=\vert Z\vert_{fix}^2 =\vert
G_{0000}\vert_{fix}^2.\label{gravitino}\eeq\noindent This formula
is to be compared with the ones of Eqs.
(\ref{centralcayley})-(\ref{entropy}) used in the black hole
context. Clearly the gravitino mass squared in the flux
compactification scenario seems to be an analogous quantity to the
black hole entropy\cite{OConnell}.

What is the physical meaning of the remaining nonzero amplitudes?
It is easy to see that they are featuring the complex mass matrix
of chiral fermions defined in an arbitrary point in moduli space.
This quantity is defined as\cite{Kalloshnew} \beq
M_{\hat{A}\hat{B}}\equiv
D_{\hat{A}}D_{\hat{B}}Z.\label{massdefi}\eeq\noindent At the
attractor point this matrix becomes a certain function of the
fluxes i.e. $(M_{\hat{A}\hat{B}})_{fix}$. After splitting the flat
indices as $\hat{A}=(\hat{I},\hat{4})$ with $\hat{I}=1,2,3$ one
can show\cite{Kalloshnew} that \beq
D_{\hat{I}}D_{\hat{J}}Z=C_{\hat{I}\hat{J}\hat{K}}\overline{D}^{\hat{4}}\overline{D}^{\hat{K}}
\overline{Z}=C_{\hat{I}\hat{J}\hat{K}}\overline{M}^{\hat{4}\hat{K}},\label{szokasoskomponensek}\eeq\noindent
where $M_{\hat{4}\hat{I}}$ is the mass matrix of the
axino-dilatino mixing with the complex structure modulino. From
Eqs.(\ref{douglas}) and (\ref{4qbitform}) it is obvious that
$M_{\hat{4}\hat{1}}=G_{1001}$, $M_{\hat{4}\hat{2}}=G_{1010}$ and
$M_{\hat{4}\hat{3}}=G_{1100}$, hence after using
$M_{\hat{4}\hat{4}}=0$ the final form of $M_{\hat{A}\hat{B}}$ is
\beq M_{\hat{A}{\hat
B}}=\begin{pmatrix}0&G_{0011}&G_{0101}&G_{1001}\\
G_{0011}&0&G_{0110}&G_{1010}\\
G_{0101}&G_{0110}&0&G_{1100}\\
G_{1001}&G_{1010}&G_{1100}&0\end{pmatrix}.\label{mass}\eeq\noindent
The explicit form of the matrix $(M_{\hat{A}\hat{B}})_{fix}$ is
given by using the expressions as given by Eq.(\ref{amplitudok}).

 Let us finally comment on the structure of the $SL(2)^{\times 4}$
invariants for our model. As the algebraically independent
$SL(2)^{\times 4}$ invariants\cite{Luque} one can take the
quantities of Ref.\cite{Lev4} with explicit expressions\beq
I_1=-\frac{1}{4}(a^2+b^2+c^2+d^2),\qquad
I_2=\frac{1}{6}(ab+ac+ad+bc+bd+cd),\label{4invek1}\eeq\noindent
\beq I_3=-\frac{1}{4}(abc+acd+bcd+abd),\qquad
I_4=abcd,\label{4invek2}\eeq\noindent where \beq a=i(t+z),\qquad
b=i(t-z),\qquad c=i(y-x),\qquad
d=i(y+x).\label{szotar}\eeq\noindent

With these notations it easy to check that our "attractor state"
$\vert G\rangle_{fix}$ is of the form
\begin{eqnarray}\vert G\rangle_{fix}&=&\frac{1}{2}(a+d)\left(\vert
0000\rangle-\vert 1111\rangle\right)+\frac{1}{2}(a-d)\left(\vert
0011\rangle - \vert 1100\rangle\right)\\\nonumber
&+&\frac{1}{2}(b+c)\left(\vert 0101\rangle-\vert
1010\rangle\right)+\frac{1}{2}(b-c)\left(\vert 0110\rangle -\vert
1001\rangle \right).\end{eqnarray}\noindent This state up to some
phase conventions is of the same form as the generic class of four
qubit entangled states\cite{Verstraete}. The state $\vert
G\rangle_{fix}$ is the result of a distillation procedure similar
in character to the one discussed in the black hole context. In
the literature this state is tackled on the same footing as the
famous GHZ state in the three-qubit case of maximal multipartite
entanglement . However, as far as the fine details of entanglement
properties are concerned there are notable differences between the
attractor state of Eq.(\ref{finalstate}) with e.g. $
p^0=q_1=q_2=q_3=0$ of GHZ type and $\vert G\rangle_{fix}$ (see
Ref.\cite{Verstraete} for more details).

Let us calculate the norm squared $\vert\vert G\vert\vert^2$ of our state
$\vert G\rangle$ at the attractor point.
One half this norm squared is an analogous quantity to the black hole potential
of Eq.(\ref{blackholepotential}) (see also Eq.(\ref{BHpot1})). Being a quantity depending merely on the fluxes at the attractor point it should be an $SL(2)^{\times 4}$
i.e. a four-qubit invariant.
For our example this quantity is also related to the sum of the a
gravitino and chiral fermion mass squares.
A quick calculation shows that 
\beq
\frac{1}{2}\vert\vert G\vert\vert^2_{fix}= 2I_1=\int F_3\wedge H_3 ,\eeq\noindent
hence the invariant we get is the standard symplectic invariant specialized to our four-qubit case.

Another interesting quantity to look at in our flux compactification
example is the four-qubit generalization of Cayley's
hyperdeterminant\cite{Cayley} known from Eq.(\ref{entropy}).
 For
the definition of this $SL(2)^{\times 4}$ and permutation
invariant polynomial of order $24$ we refer to the
literature\cite{Luque,Lev4} here we merely give its explicit form
for our example \beq D_{4}=
(-Q_0P^1P^2P^3)(p^0q_1q_2q_3)\prod_{{lkji}\in({\mathbb
Z}_2)^{\times
4}}\left((-1)^lt+(-1)^kz+(-1)^jy+(-1)^ix\right).\label{hyper4}\eeq\noindent
It is easy to check that $D_4>0$ due to our sign conventions of
Eq.(\ref{signconve}). A necessary condition for $D_4\neq 0$ for
this example of $8$ nonvanishing fluxes is the nonvanishing of the
$4$ independent amplitudes of $\vert G\rangle_{fix}$ showing up in
the $16$ terms of the product.

\section{Fermionic entanglement from toroidal compactification}

\subsection{An interpretation via fermionic systems}
As a generalization of our considerations giving rise to qubits
now we go one step further and consider the problem of obtaining
entangled systems of more general kind from toroidal
compactification. The trick is to embed our simple systems
featuring few qubits into larger ones. Here we discuss the natural
generalization of embedding qubits (based on entangled systems
with distinguishable constituents) into fermionic systems (based
on entangled systems with indistinguishable ones\cite{eckert,ghirardi}). In the quantum
information theoretic context this possibility has already been
elaborated\cite{VL}, here we show that toroidal compactifications
also incorporate this idea quite naturally.

In order to elaborate on this problem we recall the illustrative
example of Moore\cite{Moore} discussing the structure of attractor
varieties for $IIB/T^6$. As in the special case of the stu model
we choose analytic coordinates for the complex torus such that the
holomorphic one-forms are defined as $dz^a=du^a+\tau^{ab}dv^b$
where now $\tau^{ab},\quad 0\leq a,b\leq 3$ is the period matrix
of the torus with the convention \beq
{\tau}^{ab}=x^{ab}-iy^{ab}.\label{matrconvention}\eeq\noindent For
principally polarized Abelian verieties we have the additional
constraints \beq {\tau}^{ab}={\tau}^{ba},\qquad
{y}^{ab}>0.\label{matrconstraints}\eeq\noindent
 We choose as usual ${\Omega}_0=dz^1\wedge dz^2\wedge dz^3$, and
the orientation $\int_{T^6} du^1\wedge dv^1\wedge du^2\wedge
dv^2\wedge du^3\wedge dv^3 =1$.

Unlike in our considerations of the stu model now we exploit the
full $20$ dimensional space of $H^3(T^6,{\mathbb C})$. We expand
$\Gamma\in H^3(T^6,{\mathbb C})$ in the basis similar to
Eqs.(\ref{nullasok})-(\ref{egyesek}) satisfying
$\int_{T^6}\alpha^I\wedge \beta_J=\delta^I_J,\qquad I,J=1,2,\dots
10$, \beq \alpha_0=du^1\wedge du^2\wedge du^3,\qquad
\alpha_{ab}=\frac{1}{2}{\varepsilon}_{aa^{\prime}b^{\prime}}du^{a^{\prime}}\wedge
du^{b^{\prime}}\wedge dv^b \eeq \noindent \beq \beta^0=-dv^1\wedge
dv^2\wedge dv^3,\qquad
\beta^{ab}=\frac{1}{2}{\varepsilon}_{ba^{\prime}b^{\prime}}du^a\wedge
dv^{a^{\prime}}\wedge dv^{b^{\prime}}. \eeq\noindent  One can then
show that \beq
\Omega_0=\alpha_0+{\tau}^{ab}\alpha_{ab}+{\tau^{\sharp}}_{ab}\beta^{ba}-({\rm
Det}\tau)\beta^0,\label{kifejtes2}\eeq\noindent where
${\tau}^{\sharp}$ is the transposed cofactor matrix satisfying
$\tau\tau^{\sharp}={\rm Det}(\tau) I$, where $I$ is the $3\times
3$ identity matrix. Using Eq.(\ref{kifejtes2}), the usual
expression of Eq.(\ref{Kpot}) and the identity \beq {\rm
Det}(A+B)={\rm Det}A+{\rm Det}B+{\rm
Tr}(A^{\sharp}B+AB^{\sharp}),\label{detsum} \eeq\noindent valid
for $3\times 3$ matrices over ${\mathbb C}$ one can check that
\beq e^{-K}=8{\rm Det}y.\label{matrixkahler}\eeq\noindent

An element $\Gamma$ of $H^2(T^6,{\mathbb C})$ can be expanded as
\beq
\Gamma=p^0\alpha_0+P^{ab}\alpha_{ab}-Q_{ab}\beta^{ab}-q_0\beta^0.\eeq\noindent
We can rewrite this as \beq \Gamma=\sum_{1\leq A<B<C\leq
6}\gamma_{ABC}f^A\wedge f^B\wedge f^C
\label{fermistate}\eeq\noindent where \beq
(f^1,f^2,f^3,f^4,f^5,f^6)\equiv
(f^1,f^2,f^3,f^{\overline{1}},f^{\overline{2}},f^{\overline{3}})=(du^1,du^2,du^3,dv^1,dv^2,dv^3).\eeq\noindent
Here ${\gamma}_{ABC}$ is a completely antisymmetric tensor of rank
three with $20$ independent components. In the context of BPS and non-BPS
attractors clearly we have $\Gamma\in H^2(T^6,{\mathbb Z})$ with
the components $\gamma_{ABC}$ identified with the $20$ charges
$(p^0,P^{ab} ,Q_{ab},q_0)$. This identification is given by the
explicit expressions \beq p^0=\gamma_{123},\qquad
\begin{pmatrix}P^{11}&P^{12}&P^{13}\\
               P^{21}&P^{22}&P^{23}\\
           P^{31}&P^{32}&P^{33}\end{pmatrix}
=\begin{pmatrix}\gamma_{23\overline{1}}&\gamma_{23\overline{2}}&\gamma_{23\overline{3}}\\
\gamma_{31\overline{1}}&\gamma_{31\overline{2}}&\gamma_{31\overline{3}}\\
\gamma_{12\overline{1}}&\gamma_{12\overline{2}}&\gamma_{12\overline{3}}\end{pmatrix},
\label{pma}\eeq\noindent

\beq q^0=\gamma_{\overline{1}\overline{2}\overline{3}},\qquad
\begin{pmatrix}Q_{11}&Q_{12}&Q_{13}\\
               Q_{21}&Q_{22}&Q_{23}\\
               Q_{31}&Q_{32}&Q_{33}\end{pmatrix}
=-\begin{pmatrix} \gamma_{1\overline{2}\overline{3}}&
\gamma_{1\overline{3}\overline{1}}&
\gamma_{1\overline{1}\overline{2}}\\
\gamma_{2\overline{2}\overline{3}}&
\gamma_{2\overline{3}\overline{1}}&
\gamma_{2\overline{1}\overline{2}}\\
\gamma_{3\overline{2}\overline{3}}&
\gamma_{3\overline{3}\overline{1}}&
\gamma_{3\overline{1}\overline{2}}\end{pmatrix}.\label{qma}\eeq\noindent
Using the language of fermionic entanglement\cite{eckert,ghirardi}
$\Gamma$ can also be regarded as an unnormalized three fermion
state with six    single particle states\cite{VL}.

Now we introduce the new moduli dependent basis vectors \beq
e^A=f^{A^{\prime}}{S_{A^{\prime}}}^A, \qquad
{S_{A^{\prime}}}^A=\begin{pmatrix}I&I\\\tau&\overline{\tau}.\end{pmatrix}
\eeq\noindent One can then write \beq
\Gamma=\frac{1}{6!}\Gamma_{A^{\prime}B^{\prime}C^{\prime}}
\left(-ie^{K/2}e^{A^{\prime}} \wedge e^{B^{\prime}}\wedge
e^{C^{\prime}}\right), \label{indish} \eeq \noindent where \beq
\Gamma_{A^{\prime}B^{\prime}C^{\prime}}={{\cal S}_{A^{\prime}}}^A
{{\cal S}_{B^{\prime}}}^B{{\cal S}_{C^{\prime}}}^C\gamma_{ABC},
\label{kisgamma2} \eeq \noindent and \beq {\cal S}\equiv
-ie^{-K/6}
S^{-1}=-ie^{-K/6}(\tau-\overline{\tau})^{-1}\begin{pmatrix}-
\overline{\tau}&I\\\tau&-I\end{pmatrix}.\eeq\noindent In this new
form the amplitudes $\Gamma_{ABC}$ are depending on the charges
and the moduli. Notice also that now we have the {\it same} matrix
${\cal S}\in GL(6,{\mathbb C})$ acting on all indices of
$\gamma_{ABC}$. This reflects the well-known fact known from the
theory of quantum entanglement that the SLOCC
group\cite{Dur,eckert} for a quantum system consisting of {\it
indistinguishable} subsystems (now with six single particle
states\cite{VL}) is represented by the same $GL(6,{\mathbb C})$
matrices acting on each entry of a tensor representing the set of
amplitudes (now of a tripartite system). Notice also that the
basis states $-ie^{K/2}e^A\wedge e^B\wedge e^C$ for $1\leq
A<B<C\leq 6$ now form an orthonormal basis.

It is instructive to see how do we recover the stu case studied in Section III.
In particular one would like to see how the {\it indistinguishable} character of the subsystems represented by $\Gamma$
of Eq, (\ref{indish}) boils down to the {\it distinguishable} one of the subsystems represented by Eq. (\ref{statecohomology}).
In order to see this just notice that in the stu case we have merely $8$ nonzero amplitudes to be used in Eq. (\ref{kisgamma2}). Namely  we have $\gamma_{ABC}$ with
labels ${123},{12\overline{3}},{1\overline{2}3},\dots,{\overline{1}\overline{2}3},{\overline{1}\overline{2}\overline{3}}$.
Moreover, the $3\times 3$ matrix $\tau$ is now diagonal, hence the explicit form of ${\cal S}$ is
\beq
{\cal S}=\frac{1}{2}e^{-K/6}\begin{pmatrix}-\overline{\tau}^1/y^1&0&0&1/y^1&0&0\\0&-\overline{\tau}^2/y^2&0&0&1/y^2&0\\
0&0&-\overline{\tau}^3/y^3&0&0&1/y^3\\
\tau^1/y^1&0&0&-1/y^1&0&0\\
0&\tau^2/y^2&0&0&-1/y^2&0\\
0&0&\tau^3/y^3&0&0&-1/y^3\end{pmatrix}
\label{kiirva}
\eeq
\noindent
After switching to our usual ordering convention let us make the correspondence ${321}\leftrightarrow {000}, {32\overline{1}}\leftrightarrow {001}$ etc. meaning that the labels $1$ and $\overline{1}$, $2$ and $\overline{2}$, $\dots$ refer to the labels $0$ and $1$ of the {\it first }, {\it second} $\dots$ qubit.
Looking at the structure of the tensor product ${\cal S}\otimes {\cal S}\otimes {\cal S}$ and recalling that $e^{-K/2}=\sqrt{8y^1y^2y^3}$ we quickly recover the structure of the matrices of Eq.(\ref{matricak}).
According to Ref.\cite{VL} this embedding of a system with distinguishable constituents to a larger  fermionic one with indistinguishable ones is a useful trick for studying the entanglement properties of simple embedded systems.

\subsection{BPS attractors}

Let us now consider the structure of BPS attractors\cite{Moore} in
our entanglement based approach. Similar to the stu case the
attractor equations are demanding that only the $H^{3,0}$ and
$H^{0,3}$ parts of the cohomology classes are nonvanishing. This
implies that for our "state" of fermionic entanglement at the
horizon we have \beq \Gamma_{\rm
fix}=\Gamma_{321}(-ie^{K/2}e^3\wedge e^2\wedge e^1)_{\rm{fix}}-
\Gamma_{\overline{321}}(-ie^{K/2}e^{\overline{3}}\wedge
e^{\overline{2}}\wedge e^{\overline{1}})_{\rm fix}, \eeq\noindent
where \beq\overline{\Gamma}_{321}=
-\Gamma_{\overline{3}\overline{2}\overline{1}}= Z(\tau_{\rm
fix},p^0,q_0,P,Q). \label{attractorfermi} \eeq \noindent According
to the general theory\cite{VL} for classifying the SLOCC
entanglement types\cite{Dur} for tripartite fermionic systems with
six single particle states, such attractor states belong to the
fermionic generalization of the usual GHZ state well-known for
three qubits. Hence our result on the reinterpretation of the
attractor mechanism as a quantum information theoretic
distillation procedure in this fermionic context still holds.

Let us now find the explicit form of our attractor state of
Eq.(\ref{attractorfermi}). In order to do this we just have to
recall the steps discussed by Moore\cite{Moore} within the realm
of our entanglement based context. Clearly the attractor equations
now correspond to the usual ones stating that except for the ones
${\Gamma}_{321}$ and $\Gamma_{\overline{321}}$ all the amplitudes
of the fermionic sate $\Gamma$ of Eq.(\ref{indish}) vanish at the
black hole horizon. An equivalent form of this constraint can be
shown to be\cite{Moore} \beq {\rm Im}(2\overline{C})=p^0,\qquad
{\rm Im}(2\overline{C}\tau)=P, \label{12e} \eeq \noindent \beq
{\rm Im}(2\overline{C}{\rm Det}\tau)=q_0,\qquad {\rm
Im}(2\overline{C}\tau^{\sharp})=-Q, \label{34e} \eeq \noindent
where in these equations \beq C=e^{K/2} Z(\tau_{\rm fix}
,p^0,q_0,P,Q),\qquad \tau=\tau_{\rm fix}\equiv\tau(p^0,q_0,P,Q).
\label{c} \eeq \noindent For completeness let us revisit the main
steps of the derivation of the stabilized values of $\tau$ as
presented in Ref.\cite{Moore} The first equation of Eq.(\ref{12e})
can be solved by the ansatz \beq 2\overline{C}=\xi_0+ip^0.
\label{cbar} \eeq \noindent Plugging this into the second one of
Eq.(\ref{12e}) one obtains \beq p^0\tau=\lambda{\cal Y}+P,\qquad
{\rm where}\qquad \lambda\equiv\frac{C}{\vert C\vert},\quad {\cal
Y}\equiv 2\vert C\vert y. \label{taukifej} \eeq \noindent Using
this we get \beq
(p^0\tau)^{\sharp}=(p^0)^2\tau^{\sharp}=\lambda^2{\cal
Y}^{\sharp}+P^{\sharp}+\lambda{\cal Y}\times P, \label{linearno}
\eeq \noindent where ${\cal Y}\times P$ is the linearization of
the sharp map\cite{krut}. We would like to use this expression in
the second of Eq.(\ref{34e}). Since $2\overline{C}\lambda$ is a
real number, the term linear in $\lambda$ does not give
contribution to the terms under ${\rm Im}(\dots)$. As a
consequence of this one gets \beq {\cal
Y}^{\sharp}=p^0Q+P^{\sharp}. \label{calyexpress} \eeq \noindent
Readers familiar with the theory of Freudenthal triple
systems\cite{krut} realize this expression as one of the {\it
quadratic rank polynomials}. The Freudenthal triple system now in
question is the one based on the cubic Jordan algebra of $3\times
3$ complex matrices. Such polynomials are needed for the
classification of Freudenthal systems, which in turn has also
relevance to classification of entanglement types\cite{VL,Bor1} of
special quantum systems with few subsystems. Notice also that a
necessary condition for nondegenerate tori is \beq {\rm Det}{\cal
Y}^{\sharp}=({\rm Det}({\cal Y}))^2\neq
0\label{kihasznal},\eeq\noindent moreover for a polarized Abelian
variety ${\cal Y}^{\sharp}$ is a symmetric positive matrix
.(Recall the third expression in Eq.(\ref{taukifej}) taken
together with ${\cal Y}^{\sharp}{\cal Y}=({\rm Det}{\cal Y})I$,
and $e^{-K}=8{\rm Det y}>0, y>0.$)

Using Eq.(\ref{calyexpress}) in the first of Eq.(\ref{taukifej})
provides an expression for $\tau$ in terms of the charges {\it
and} the unknown quantity $C=e^{K/2}Z$. In order to determine its
value in terms of the charges  we now turn to the first equation
of Eq.(\ref{34e}). First we use the identity of Eq.(\ref{detsum})
 to get
\beq {\rm Det}(p^0\tau)=(p^0)^3{\rm Det}\tau={\lambda}^3{\rm
Det}{\cal Y}+{\lambda}^2(p^0{\rm Tr}(PQ)+3{\rm
Det}(P))+\lambda{\rm Tr}({\cal Y}P^{\sharp})+{\rm Det}(P).
\label{reszlet} \eeq \noindent Using this in the first of
Eq.(\ref{34e}) after some manipulations one obtains \beq
\frac{\xi_0}{\vert C\vert}{\rm Det}{\cal Y}=-\tilde{p}^0,\qquad
\tilde{p^0}\equiv 2{\rm Det}P+p^0({\rm Tr}(PQ)+p^0q_0).
\label{vegkifejlet} \eeq \noindent The expression on the right
hand side which is cubic in the charges is also a well-known
quantity in the theory of Freudenthal triple systems. It is a part
of the Freudenthal dual charge configuration\cite{Bor1}
$(\tilde{p}^0,\tilde{q}_0,\tilde{P}^{ab},\tilde{Q}_{ab})$ (which
is also used as one of the amplitudes of the dual entangled state
in Ref.\cite{VL}) based on a trilinear operator\cite{krut}.

Let us now take the square of the first equation of
Eq.(\ref{vegkifejlet}) and express $({\rm Det}{\cal Y})^2$ using
Eq.(\ref{kihasznal}) in terms of ${\rm Det}{\cal Y}^{\sharp}= {\rm
Det}(p^0Q+P^{\sharp})$. The determinant of the sum of matrices can
be tackled again by Eq.(\ref{detsum}) yielding the result \beq
\xi_0^2=\frac{(\tilde{p}^0)^2}{\cal D},\label{xiveg}\eeq\noindent
where \beq {\cal D}=-(p^0q_0+{\rm Tr}(PQ))^2+4{\rm
Tr}(P^{\sharp}Q^{\sharp})+4p^0{\rm Det}Q-4q_0{\rm
Det}P.\label{cartan}\eeq\noindent Note that ${\cal D}$ is minus
half of the usual quartic invariant of Freudenthal triple
systems\cite{krut}.

For BPS solutions we chose the branch \beq
\xi_0=-\frac{\tilde{p}^0}{\sqrt{\cal D}},\label{xiuj}\eeq\noindent
provided ${\cal D}>0$. Comparing this with the first of
Eq.(\ref{vegkifejlet}) one gets ${\rm Det}{\cal Y}=\vert C\vert
\sqrt{\cal D}$. Using this and Eq.(\ref{calyexpress}) with the
third of Eq.(\ref{taukifej}) one gets \beq y=\frac{1}{2}\sqrt{\cal
D}(p^0Q+P^{\sharp})^{-1}.\label{kepzetes}\eeq\noindent Similar
manipulations using the first of Eq.(\ref{taukifej}) yield for the
real part of $\tau$ \beq x=\frac{1}{2}(2PQ-[p^0q_0+{\rm
Tr}(PQ)]I)(p^0Q+P^{\sharp})^{-1}.\label{valos}\eeq\noindent One
can check that the stu case of Eq.(\ref{ittaveg}) is recovered
when using diagonal matrices for ${\tau}=x-iy$, $P$ and $Q$.

Using these results one can show that the GHZ-like state at the
horizon, as the result of a distillation procedure, is of the form
as given by Eq.(\ref{finalstate}) with suitable replacements.
First Cayley's hyperdeterminant ${\cal D}$ has to be replaced by
its generalization $D$ as given by Eq.(\ref{cartan}). Moreover,
the phase $\alpha$ of the central charge is determined by the
equation \beq
\tan\alpha=\sqrt{-D}\frac{p^0}{\tilde{p}^0},\label{ujfazis}\eeq\noindent
where $\tilde{p}^0$ is given by the quantity showing up in
Eq.(\ref{vegkifejlet}). The stabilized  states $\vert
000\rangle_{fix}$ and $\vert 111\rangle_{fix}$ of
Eq.(\ref{finalstate}) should be replaced by their "fermionic"
counterparts $(-ie^{K/2} e^3\wedge e^2\wedge e^1)_{\rm{fix}}$ and
$(-ie^{K/2}e^{\overline{3}}\wedge e^{\overline{2}}\wedge
e^{\overline{1}})_{\rm fix}$.

For BPS black holes we have $M^2=\vert Z\vert^2$ hence the
Bekenstein-Hawking entropy of the extremal, spherically symmetric
black hole is $S_{BH}=\pi M^2=\pi\vert Z\vert^2$. Since
$C=e^{K/2}Z$ and ${\rm Det}Y=8\vert C\vert^3{\rm Det}y=\vert
C\vert^3e^{-K}=\vert C\vert\sqrt{\cal D}$ one gets for the entropy
\beq S_{BH}=\pi \sqrt{{\cal
D}(\Gamma)},\label{entropyfermionic}\eeq\noindent with ${\cal D}$
is given by Eq.(\ref{cartan}).

Based on our experience with the STU case where the entropy
formula was given in terms of a genuine tripartite measure (i.e.
$\tau_{123}\equiv 4\vert D\vert$ i.e. the
three-tangle\cite{Kundu}), it is tempting to interpret ${\cal
T}_{123}\equiv 4\vert{\cal D}\vert$ as an entanglement measure for
three fermions with six single particle states as represented by
the state $\Gamma$ Eq.(\ref{fermistate}). (The extra factor of $4$
is only needed for normalized states in order to restrict the
values of this entanglement measure to the interval $[0,1]$.)
According to Ref.\cite{VL} within the realm of quantum information
the quantity $4\vert{\cal D}\vert$ indeed works as a basic
quantity to characterize the entanglement types under the SLOCC
group\cite{Dur}. Within the context of black hole solutions we
know that the unnormalized states in question are either charge
states with integer amplitudes or ones satisfying extra reality
conditions, hence the SLOCC group should be restricted to its
suitable real subgroup i.e. the U-duality group. Based on the
results of Ref.\cite{VL} it is not difficult to see that the
different types of black holes should correspond to the different
entanglement types of fermionic entanglement. This correspondence
runs in parallel with the observation of Kallosh and
Linde\cite{KL} that the entanglement types of three qubit states
correspond to different types of stu black holes.

\section{Conclusions}

In this paper we have shown how  qubits are
arising
from the geometry of tori serving as extra dimension in IIB compactifications.
Our results clarified some of the issues left unclear in the paper of Borsten et.al.\cite{Borsten}
In particular the investigations of that paper interpreting wrapped branes as qubits were lacking an explicit construction of the Hilbert space where these qubits live.
Here we have identified this space inside the cohomology of tori.
Moreover, we have also shown that the Hodge diagonal basis usually used in the supergravity literature is naturally connected to the charge and moduli dependent
multiqubit states used in our recent papers\cite{Levay1,Levay4qbit,Levayattr}.
This result provides the simplest way to understand the well-known attractor mechanism as a distillation process an issue elaborated in our previous set of papers. 
The idea "qubits from extra dimensions" have also turned out to be very useful to 
generalize the black hole-qubit correspondence to some sort of flux-attractor-qubit correspondence. Indeed, for toroidal models it is quite natural to extend our considerations to new attractors of that kind\cite{Kalloshnew,Denefdouglas}.
We pointed out that four-qubit systems are characterizing some of the key issues for such models\cite{OConnell}.

Though our main motivation was to account for the occurrence of qubits 
in these exotic scenarios we have revealed that
in the string theoretical  context
 entangled systems of more general kind than qubits should rather be considered.
In particular for toroidal models we have seen that the natural arena where these systems  live is
 the realm of fermionic entanglement\cite{eckert,ghirardi} of
 subsystems with {\it indistinguishable} parts. The notion "fermionic"
 entanglement is simply associated with the structure of the
 cohomology of $p$-forms related to $p$-branes. 
As it has already been pointed out in our recent paper on special entangled systems\cite{VL}, qubits are  arising as embedded systems with distinguishable constituents inside such fermionic ones.
Interestingly compactification on $T^6$ in the IIB duality frame\cite{Moore} provides a particularly nice manifestation of this idea.

Notice that in our examples of toroidal compactification we merely discussed BPS black holes. However, the attractor mechanism as a distillation procedure also works for non-BPS attractors\cite{Levayattr}.
For the STU model it turns out that for the non-BPS branch $\vert\Gamma\rangle_{fix}$ will be again in the GHZ class where now none of the amplitudes are vanishing, however their magnitudes are equal. The relative signs of these amplitudes can be characterized via an error correction framework\cite{Levayattr} based on the flat covariant derivarives acting as projective bit flips as shown in Sections II. and III.

Why only tori?
Clearly we should be able to remove the rather disturbing restriction to toroidal compactifications by embarking on the rich field of Calabi-Yau compactifications. Notice in this respect that the decompositions of Eqs. (\ref{expand}) and (\ref{douglas}) in the Hodge diagonal basis can be used to reinterpret such formulas as
	{\it qudits} i.e. $d$-level systems with $d=h^{2,1}+1$ in the type IIB duality frame.
F-theoretical flux compactifications for elliptically fibered Calabi-Yau fourfolds can then be associated with
entangled systems comprising a qubit (a $T^2$ accounting for the axion-dilaton) and a qudit coming from a Calabi-Yau three-fold ($CY_3$).
Alternatively after using instead of $CY_3$ the combination $T^2\times K3$
we can have tripartite systems consisting of two qubits and a qudit etc.
The idea that separable states geometrically should correspond to product manifolds and entangled ones to fibered ones was already discussed in the literature, for the simplest cases of two and three qubits\cite{Levmetric}.
It would be interesting to explore further consequences of this idea in 
connection with the black hole-flux attractor-qubit correspondence.

\section{Acknowledgement}
The author would like to thank Professor Werner Scheid for the
warm hospitality at the Department of Theoretical Physics of the
Justus Liebig University of Giessen where part of this work has
been completed. This work was supported by the New Hungary
Development Plan (Project ID: T\'AMOP-4.2.1/B-09/1/KMR-2010-002),
and the DFG-MTA project under contract No.436UNG113/201/0-1.


\begin{thebibliography}{}
\bibitem{Borsten} L. Borsten, D. Dahanayake, M. J. Duff, H.
Ebrahim and W. Rubens, Phys. Rev. Lett. {\bf 100}, 251602 (2008).
\bibitem{Duff} M. J. Duff, Phys. Rev. D{\bf 76}, 025017 (2007).
\bibitem{KL} R. Kallosh and A. Linde, Phys. Rev. D{\bf 73} 104033
(2006).
\bibitem{Levay1} P. L\'evay, Phys.Rev. D{\bf 74}, 024030 (2006).
\bibitem{review} L. Borsten, D. Dahanayake, M. J. Duff, H.
Ebrahim, W. Rubens, Physics Reports, {\bf 471}, 113 (2009).
\bibitem{e7}M. J. Duff and S. Ferrara, Phys. Rev. D{\bf 76},
124023 (2007), P. L\'evay, Phys. Rev. D{\bf 75}, 024024 (2007), M.
J. Duff and S. Ferrara, Phys. Rev. D{\bf 76}, 124023 (2007), P.
L\'evay, M. Saniga, P. Vrana and P. Pracna, Phys. Rev. D{\bf 79},
084036 (2009).
\bibitem{Verstraete} F. Verstraete, J. Dehaene, B. DeMoor, H. Verscheide,
  Phys. Rev. A{\bf 65}, 052112
(2002).
\bibitem{Levay4qbit} P. L\'evay, Phys. Rev. D{\bf 82}, 026002
(2010).
\bibitem{Duff4qbit} L. Borsten, D. Dahanayake, M. J. Duff, A.
Marrani, W. Rubens, Physical Review Letters {\bf 105}, 100507
(2010), L. Borsten, M. J. Duff, A. Marrani, W. Rubens, Eur. Phys.
J.Plus, {\bf 126} 37 (2011).
\bibitem{Dur} W. D\"ur, G. Vidal, and J. I. Cirac, Phys. Rev. {\bf A62}, 062314 (2000).
\bibitem{3qbitborsten} L. Borsten, D. Dahanayake, M. J. Duff, H.
Ebrahim, W. Rubens, Phys. Rev. A{\bf 80}, 032326 (2009).
\bibitem{elliplevay} P. L\'evay, Phys. Rev. D{\bf 84}, 025023 (2011).
\bibitem{ferrorbit} S. Ferrara, A. Marrani, E. Orazi, R. Stora, A.
Yerayan, Journal of Mathematical Physics {\bf 52}, 062302 (2011).
\bibitem{VL} P. L\'evay and P. Vrana, Phys. Rev. {\bf A}78, 022329
(2008), P. Vrana and P. L\'evay J. Phys. A: Math. Theor. {\bf 42},
285303 (2009).
\bibitem{attractors}
S. Ferrara, R. Kallosh and A. Strominger, Phys. Rev. D{\bf 52},
5412 (1995), A. Strominger, Phys. Lett. B{\bf 383}, 39 (1996), S.
Ferrara and R. Kallosh, Phys. Rev. D{\bf 54}, 1514 (1996), S.
Ferrara and R. Kallosh, Phys. Rev. D{\bf 54}, 1525 (1996), K.
Goldstein, N. Izuka, R. P. Jena and S. P. Trivedi, Phys. Rev.
D{\bf 72} 124021 (2005), A. Sen, Journal of High energy Physics
{\bf 0509}, 038 (2005), P. K. Tripathy and S. Trivedi, Journal of
High Energy Physics {\bf 0603} 022 (2006).
\bibitem{Levayattr} P. L\'evay, Phys. Rev. D{\bf 76}, 106011 (2007), P. L\'evay
and Sz. Szalay, Phys. Rev. D{\bf 82}, 026002 (2010), P. L\'evay
and Sz. Szalay, Phys. Rev. D{\bf 83}, 045005 (2011).
\bibitem{Frascati} P. L\'evay , "Attractors, Black Holes and
Multiqubit Entanglement, The Attractor Mechanism: Proceedings of
the INFN-Laboratori Nazionali di Frascati School 2007,
Springer-Verlag 2010.
\bibitem{Behrndt} M. J. Duff, J. T. Liu, and J. Rahmfeld,
Nucl.Phys.{\bf B459}, 125 (1996), K. Behrndt, R. Kallosh, J.
Rahmfeld, M. Shmakova, and W. K. Wong, Phys. Rev. D{\bf 54}, 6293
(1996).
\bibitem{Kalloshnew} R. Kallosh, Journal of High Energy Physics
{\bf 0512}, 022 (2005).
\bibitem{OConnell} F. Larsen and R. O'Connell, Journal of High
Energy Physics, {\bf 0907} 049 (2009).
\bibitem{eckert} K. Eckert, J. Schliemann, D.Bruss and M. Lewenstein, Ann. Phys. (N.Y.) {\bf 299}, 88 (2002).
\bibitem{ghirardi} G. C. Ghirardi, L. Marinatto, Phys. Rev. A{\bf 70}, 012109
(2004).
\bibitem{Ferrarastu}
S. Bellucci, S. Ferrara, A. Marrani, A. Yeranyan, Entropy 10, 507
(2008).
\bibitem{Gimon}
E. G. Gimon, F. Larsen, J. Simon, Journal of High Energy Physics,
0801,040 (2008).
\bibitem{Suzuki} H. Suzuki, Mod. Phys. Lett. A11, 623 (1996).
\bibitem{Denef} F. Denef, Journal of High Energy Physics 0008, 050 (2000).
\bibitem{Moore} G. Moore, arXiv:hep-th/9807087
\bibitem{Lust} R. Blumenhagen, B. Körs, D. Lüst, S. Steiberger,
Physics Reports {\bf 445} 1-193 (2007).
\bibitem{Cayley} A.Cayley,
 Camb. Math. J. {\bf 4}, 193 (1845).
\bibitem{Kundu} V. Coffman, J. Kundu and W. K. Wootters, Phys.
Rev. A{\bf 61}, 052306 (2000).
\bibitem{GHZ} D. M. Greenberger, M. A. Horne, A. Zeilinger,
arXiv:0712.0921, D. Bouwmeister, J-W Pan, M. Daniell, H.
Weinfurter, A. Zeilinger, Physical Review Letters, {\bf 82}, 1345
(1999).
\bibitem{Denefdouglas} F. Denef and M. R. Douglas, Journal of High
Energy Physics, {\bf 0405}, 072 (2004).
\bibitem{Dasgupta} K. Dasgupta, G. Rajesh, and S. Sethi, Journal
of High Energy Physics, 08, 023 (1999), S. B. Giddings, S. Kachru,
and J. Polichinski, Phys. rev. D{\bf 66}, 106006, (2002).
\bibitem{Luque} G. Luque and Y.I. Thibon, Phys. Rev. A{bf 67},
042303 (2003).
\bibitem{Lev4} P. L\'evay, Journal of Physics A: Math. Gen. {\bf 39}, 9533
(1006).
\bibitem{krut} S. Krutelevich, Journal of Algebra {\bf 314}, 924, (2007).
\bibitem{Bor1} L. Borsten, D. Dahanayake, M. J. Duff, W. Rubens, Phys. Rev.
D{\bf 80}, 026003, (2009).
\bibitem{Levmetric} R. Mosseri and R. Dandoloff,
J. Phys. A: Math. Gen. {\bf 34}  10243 (2001),
B. A. Bernevig, H-D. Chen, J. Phys. A: Math. Gen. {\bf 36} 8325, (2003),
P. L\'evay, J. Phys. A: Math. Gen. {\bf 37}, 1821, (2004).

\end{thebibliography}
\end{document}